\documentclass{aa}
\usepackage{graphics,graphicx,txfonts}
\newcommand\simgt{\lower.5ex\hbox{$\;\buildrel>\over\sim\;$}}
\newcommand\simlt{\lower.5ex\hbox{$\;\buildrel<\over\sim\;$}}

\begin{document}

\title{Molecular Gas in NUclei of GAlaxies (NUGA):\\ II. The Ringed LINER NGC\,7217}
                                
\author{F. Combes \inst{1} 
\and S. Garc\'{\i}a-Burillo \inst{2}
\and F. Boone \inst{3}
\and L.K. Hunt \inst{4}
\and A.J. Baker \inst{5}
\and A. Eckart \inst{6}
\and P. Englmaier \inst{7}
\and S. Leon \inst{8}
\and R. Neri \inst{9}
\and E. Schinnerer \inst{10}
\and L.J. Tacconi \inst{5}
} 
\offprints{F. Combes, \email{francoise.combes@obspm.fr}\\
Based on observations carried out with the IRAM Plateau de Bure Interferometer and IRAM
30m telescope. IRAM is supported by INSU/CNRS (France), MPG (Germany) and IGN (Spain).} 

\institute{Observatoire de Paris, LERMA, 61 Av. de l'Observatoire, 
   F-75014, Paris, France 
\and Observatorio Astron\'omico Nacional (OAN),
    Alfonso XII, 3, 28014-Madrid, Spain 
\and Bochum University, Universit\"atstrasse 150, D-44780 Bochum, Germany
\and Istituto di Radioastronomia/CNR, Largo Enrico Fermi, 5, 50125 Firenze, Italy
\and Max-Planck-Institut f\"ur extraterrestrische Physik, Postfach 1312, D-85741 Garching, Germany
\and I. Physikalisches Institut, Universit\"at zu K\"oln, Z\"ulpicherstrasse 77, 
50937-K\"oln, Germany
\and Astronomy, Universit\"at Basel, Venusstrasse 7, CH 4102 Binningen,  Switzerland 
\and Instituto de Astrof\'{\i}sica de Andaluc\'{\i}a (CSIC), Camino Bajo de Hu\'etor, 
24, 18008 Granada, Spain
\and IRAM-Institut de Radio Astronomie Millim\'etrique, 300 Rue de la Piscine,
38406-St.Mt.d'H\`eres, FRANCE
\and  NRAO, PO Box 0, Socorro, NM-87801, USA
}
 
\date{Received XX XX, 2003; accepted XX XX, 2003}
\authorrunning{Combes et al.} 
\titlerunning{NUGA: II. NGC\,7217}

\abstract{ 
We present CO(1-0) and CO(2-1)  maps of the LINER galaxy NGC\,7217,
obtained with the IRAM interferometer, at 2.4\arcsec$\times$1.9\arcsec\ 
and 1.2\arcsec$\times$0.8\arcsec\ resolution
respectively. The nuclear ring (at r\,=\,12\arcsec\,=\,0.8kpc) dominates the CO maps, 
and has a remarkable sharp surface density gradient at its inner edge.
The latter is the site of the
stellar/H$\alpha$ ring, while 
the CO emission ring extends farther or is broader (500-600pc).
This means that the star formation has been
more intense toward the inner edge of the CO ring, in a thin layer,
just at the location of the high gas density gradient.
The CO(2-1)/CO(1-0) ratio is close to 1, typical of warm optically
thick gas with high density.
The overall morphology of the ring is quite circular, with no evidence
of non-circular velocities. In the CO(2-1) map, a central concentration
might be associated with the circumnuclear ionized gas detected inside r=3"
and interpreted as a polar ring in the literature. 
 The CO(2-1) emission inside 3" 
coincides with a spiral dust lane, 
clearly seen in the HST $V - I$ color image.\\ 
N-body simulations including gas dissipation and
star formation are performed to better understand the nature of
the nuclear ring observed. The observed rotation curve of NGC 7217
allows two possibilities, according to the adopted mass for the disk:
(1) either the disk is massive, allowing a strong bar to develop,
or (2) it is dominated in mass by an extended bulge/stellar halo, 
and supports only a mild oval distortion.
The amount of gas also plays an important role in the disk
stability, and therefore the initial gas fraction was varied, 
with star formation reducing the total gas fraction to the
observed value. The present observations support {\it only} the
bulge-dominated model, which
is able to account for the nuclear ring in CO
and its position relative to the stellar and H$\alpha$ ring.
In this model, the gas content was higher in the recent past (having been consumed
via star formation), and the structures formed were
more self-gravitating. Only a mild bar formed, which
has now vanished, but the
stars formed in the highest gas density peaks toward the inner edge
of the nuclear ring, which corresponds to the observed thin stellar ring.
We see no evidence for an ongoing fueling of the nucleus; instead,
 gas inside the ring is presently experiencing an outward flow. 
To account for the nuclear activity, some gas infall and fueling
must have occured in the recent past (a few Myr ago), 
since some, albeit very small, CO emission is detected at the very center.
These observations have been made in the context of the NUclei of GAlaxies 
(NUGA) project, aimed at the study of the different mechanisms for gas fueling of AGN.
}

\maketitle

\section{Introduction \label{intro}} 

Accretion onto black holes has become the accepted explanation for 
nuclear activity in galaxies.  
But while most galaxies are now thought to harbor central massive black
holes, only a fraction of them host active nuclei (AGNs).
The reasons for this have been the subject of much investigation, but
it remains unclear whether the explanation lies 
in the total gas mass available for fueling the AGN
or in the mechanism by which fuel is funneled into the central pc.

The main problem for fueling the AGN is the removal of angular momentum
from the disk gas, a process which can be accomplished through 
non-axisymmetric  perturbations.
These might be of external origin,
triggered by a companion (Heckman et al. 1986), or internal due to density waves 
such as bars or spirals,
and their gravity torques (e.g. Combes 2001).
The distinction is sometimes difficult to make, since the
tidal interaction of companions triggers bar formation in the
target disk.  

Nevertheless, the presence of a bar is not necessarily 
associated with AGNs (Mulchaey \& Regan 1997, 
Knapen et al 2000), and the fueling processes might involve more localised
phenomena, such as embedded nuclear bars
(Shlosman et al. 1989), lopsidedness or $m=1$ instabilities (Shu
et al. 1990; Kormendy \& Bender 1999; Garc\'{\i}a-Burillo et al.
2000) or the presence of warped nuclear disks (Pringle 1996, 
Schinnerer et al. 2000a, 2000b). 
To account for the non-correlation between bars and nuclear activity, there must
be time delays between the first gas inflow due to the bar or spiral
density waves, which first drive a nuclear starburst,
and the subsequent fueling of the black hole, for example by the
tidal disruption of stars in the just-formed nuclear stellar
clusters.
Once the gas has reached sufficiently small radii, the
dynamical friction exerted by bulge stars on the giant molecular
clouds can also provide a fueling mechanism (e.g. Stark et al. 1991).

The study of molecular gas morphology and dynamics constitutes an ideal tool for
investigating AGN fueling, and its connection with circumnuclear star formation.
Molecular gas is the dominant phase  of the interstellar medium
in the central kiloparsec of spiral
galaxies, whereas the atomic gas is deficient there. This makes CO lines
ideal to trace the dynamics of the interstellar medium and radial gas flows. 
However,
to compare all the possible mechanisms to observations, high-resolution
(1-2 arcsec) maps of the molecular component in the centers of galaxies are required.
Previous surveys of molecular gas in active galaxies have been carried out by
other groups (Heckman et al. 1989; Meixner et al. 1990;
Vila-Vilaro et al. 1998), but have had insufficient spatial resolution to resolve the
nuclear disk structures, or were limited to small samples (Tacconi et al. 1997;
Baker 2000).
This paper is the second of a series which describes on a case-by-case basis
the results of the NUGA (or NUclei of GAlaxies) project.
 A detailed description of NUGA is
 given in Garc\'{\i}a-Burillo et al. (2003a) and
Paper I (Garc\'{\i}a-Burillo et al. 2003b), which focuses on the
 counter-rotating system NGC 4826.

NGC\,7217 ($D = 14.5 Mpc$: Buta et al. 1995, hereafter B95) 
is one of the first galaxies of the sample to have been mapped, 
and this paper describes the distribution and dynamics of its molecular gas.
NGC\,7217 is an isolated LINER 2 galaxy
(Ho, Filippenko, \&  Sargent 1997) of Hubble type (R)SA(r)ab. 
Its main characteristic is its high degree of
axisymmetry, and the presence of three stellar rings,  with
radii of 0.75, 2.2 and 5.4\,kpc. All three rings are
young and bluer than the surrounding stellar disk (B95). 
The three rings are reminiscent of resonant rings in a barred galaxy
(e.g. Buta \& Combes 1996), although the 
galaxy is not barred, containing at most a weak oval perturbation (B95).
The rings however could be the remnants of a previous bar episode in this galaxy, and
it is interesting to note that one of the best correlations with nuclear activity level
is the presence of outer rings (Hunt \& Malkan 1999). Through numerical
simulations of the gas response in the potential derived from a red image of
NGC\,7217, Buta et al. (1995) showed that even the present weak oval
distortion was able to form the three observed resonant rings. 

NGC\,7217 is an early-type galaxy, and its bulge is particularly massive and extended. 
Merrifield \& Kuijken (1994) claimed that a significant fraction (30\%) of its stars are
 counter-rotating relative to the rest of the galaxy; however,
Buta et al. (1995) interpreted this instead as
a large velocity dispersion in the bulge. The bulge component is not only 
predominant in the center, but dominates the disk out to a 
large galactocentric distance. 
The disk of NGC\,7217 is rather regular, and does not seem to possess a strong
density wave; there is only flocculent multi-arm spiral structure, 
of varying prominence
over the disk (cf. Table \ref{ringtab} for the various
ring radii). 

H$\alpha$ imaging reveals a very neat and complete nuclear ring of
$\sim\,$21" diameter (Pogge 1989, Verdes-Montenegro et al. 1995), strikingly coincident
with the blue stellar nuclear ring, but not with the nuclear dust ring;
the latter is slightly interior,
probably an effect of extinction (B95).
In this paper, we compare the CO maps obtained
with the IRAM interferometer to HST/WFPC2, HST/NICMOS, and ground-based
H$\alpha$ maps, 
show that the diameters of the corresponding rings are different,
and interpret them in the light of numerical modeling. 

We perform N-body numerical simulations, including gas dissipation and star
formation, in order to illustrate the proposed mechanisms, namely
how star formation and/or dynamical
 effects can drive the evolution of the radius of the nuclear gas ring.
The previous numerical simulations of gas flow in a fixed
potential derived from a red image of the galaxy (B95) only tested 
the present disk perturbation and its pattern speed, but
did not address its formation and evolution. 
Also, the new observations presented in this paper allow us to better constrain 
the models.

The CO observations, together with images at other wavelengths,
are presented in Section 2, and the CO results in Section 3. 
Comparison with other wavelengths is performed in Section 4.
Section 5 describes the star formation history and stellar populations as
gleaned from the observations at other wavelengths, and how these relate
to the molecular gas morphology.
In Section 6 we briefly describe the code and numerical methods, followed by
the results of our simulations in Section 7.
Possible interpretations are discussed in Section 8, and Section 9 gives
our conclusions. 

\begin{table}[ht]
\caption{ Parameters for NGC\,7217 \label{param} }
\begin{flushleft}
\begin{tabular}{lll}
\hline
Parameter  &   Value                              &  Reference \\
\hline
RA (J2000)  &   22$^{\rm h}07^{\rm m}52.4^{\rm s}$ & NED \\
DEC (J2000) &   31$^\circ$21$\prime$32.2\arcsec    & center \\
 V$_{hel}$ &   952  km/s                          & B95 \\
RC3 Type   &  (R) SA (r) ab                       & deV \\
Refined type&  (R') SA (rs, nr)ab                 & B95 \\
Inclination &         36$^\circ$                  & LEDA  \\
Position angle &         95$^\circ$               & LEDA  \\
Distance    &         14.5Mpc (1\arcsec= 70pc)    & B95 \\
M(HI)                 &  0.58 10$^9$ M$_\odot$    & B95 \\
L$_B$       &    1.6 10$^{10}$ L$_{\odot}$        & LEDA \\
L$_{FIR}(40-120\,{\rm \mu m})$  & 3 10$^{9}$ L$_{\odot}$  & IRAS \\
\hline
\end{tabular}
\end{flushleft}
B95: Buta et al. (1995)\\
deV: de Vaucouleurs et al. 1991 \\
\end{table}

\begin{figure*}
\rotatebox{-90}{\includegraphics[width=17cm]{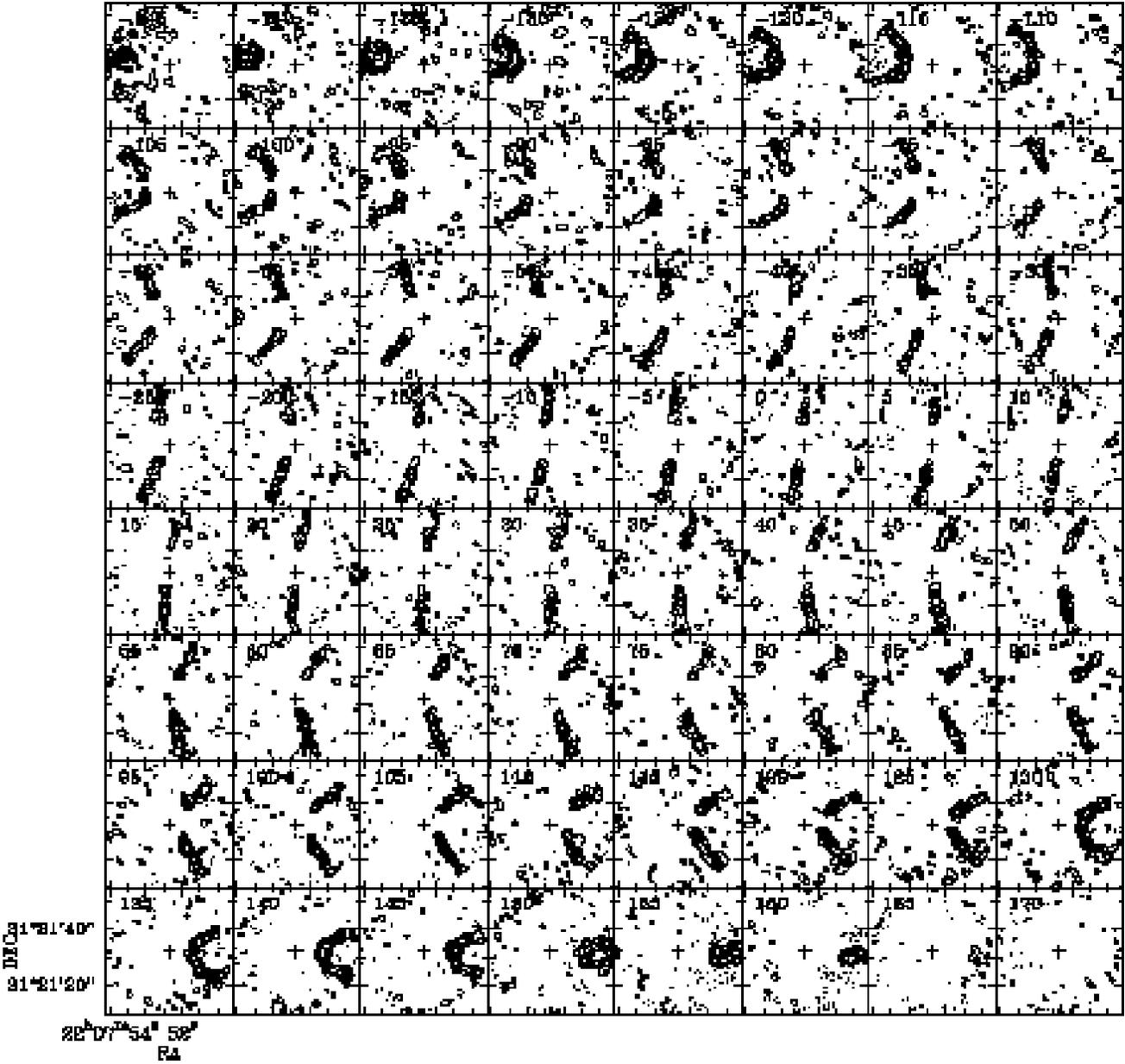}}
\caption{CO(1-0) velocity-channel maps observed with the 
IRAM interferometer in the nucleus of NGC\,7217 
with a spatial resolution of (HPBW)
 2.4$\arcsec\times$1.9$\arcsec$ (PA=48$^{\circ}$). The center
of observations, given in Table \ref{param},
is indicated by a cross at $\alpha_{J2000}= 22^{\rm h}07^{\rm m}52.4^{\rm s}$,
$\delta_{J2000}= 31^\circ$21$'$32.2$\arcsec$. Velocity-channels range from
v=-145km/s to v=170km/s in steps of $5\,{\rm km\,s^{-1}}$ relative 
to V= 952km/s LSR (or 940 km/s hel).
The maps are corrected for primary beam attenuation.
The contours begin at 10mJy/beam,
their spacing is 10mJy/beam, and the maximum is 50 mJy/beam. 
\label{chan}}
\end{figure*}

\section{Observations \label{obs}}

\subsection{IRAM single dish CO observations}

\begin{figure}[ht]
\rotatebox{-00}{\includegraphics[width=8cm]{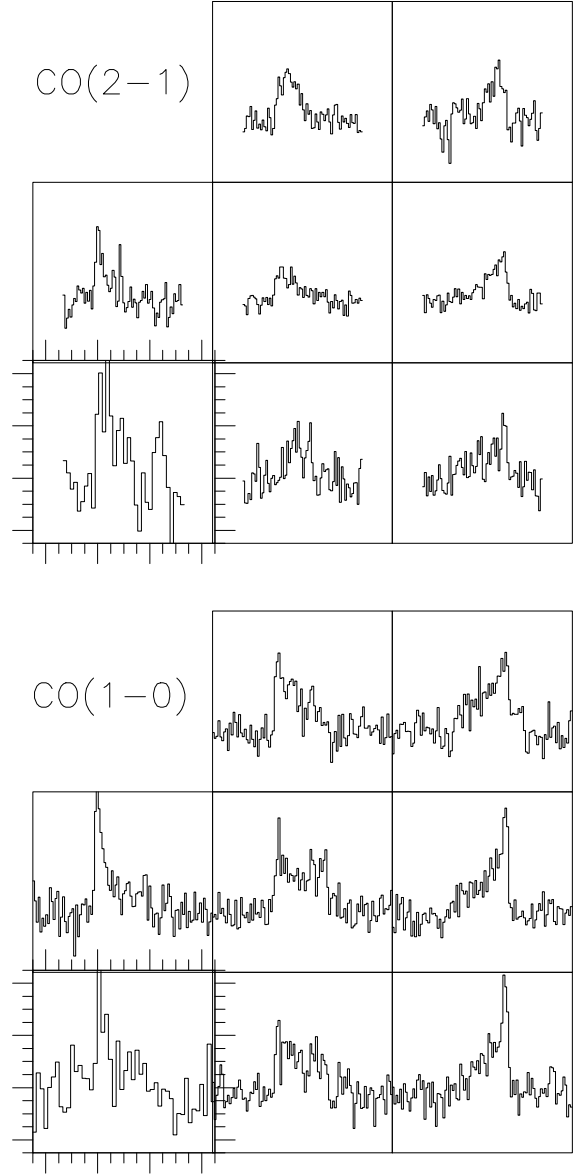}}
\caption{Small maps of NGC 7217 made with the IRAM 30m, with 7"
spacing, in CO(1-0) (bottom) and CO(2-1) (top). The velocity scale is
-500  to $500\,{\rm km\,s^{-1}}$
 relative to $V = 952\,{\rm km\,s^{-1}}$ LSR; the $T_A^*$
 temperature scale is from -.05 to 0.1K
(or T$_{mb}$ from -.063 to 0.126K for CO(1-0), and T$_{mb}$ from -.088 to 0.175K
for CO(2-1)). Lack of time prevented to complete the raster in the NE position, but
this does not significantly perturb the short-spacing correction. 
\label{30m}}
\end{figure}

To add the short spacings, we performed IRAM 30m observations
in a $3 \times 3$ raster pattern with 7" spacing
(see Fig.~\ref{30m}), in July 2002.
 We used 4 SIS receivers to observe
simultaneously at the frequencies of the CO(1--0) and the
CO(2--1) lines.  At 115 GHz and 230 GHz, the telescope
half-power beam widths are 22$''$ and 12$''$, respectively.  The
main-beam efficiency is $\eta$$_{\rm mb}=T_{\rm A}^*/T_{\rm
mb}$=0.79 at 115 GHz and 0.57 at 230 GHz.  The
typical system temperature varied between 200 and 450 K (on the $T_{\rm
A}^*$ scale) at both frequencies. Wobbler switching
mode was used, with reference
positions offset by 4$'$ in azimuth.  The pointing was regularly
checked on continuum sources and the accuracy was 3$''$ rms.  The
backends were two 1MHz filter banks and auto-correlator spectrometers.  The total
bandwidth available was 512 MHz at 115 GHz and 230 GHz, corresponding
to 1300km/s and 650 km/s for the CO(1-0) and CO(2-1) lines respectively
(with velocity resolutions of 2.6 and 1.3km/s).

Short spacings were then
included using the SHORT-SPACE task in the GILDAS
software (e.g. Guilloteau \& Lucas, 2000).
Short spacings visibilities are computed from a map built by interpolation of
the 30m observations,  deconvolved from the 30m beam and multiplied by the
PdB primary beam. The weights are adjusted in order to get the same 
mean weights in the single-dish data as in
the interferometer data in the uv range of $1.25\,D/\lambda$ to $2.5\,D/\lambda$
(D=15 m).  All figures are made with short-spacing
corrected data.

It is important to estimate the flux filtered out by the interferometric
observations. With a beam of 45\arcsec\, Young et al.
(1995) measured a CO(1-0) flux
 toward the center of I(CO)= 2.27
K.km/s (in the T$_A^*$ scale). With a conversion factor of 42 Jy/K, 
their integrated flux is then S(CO)=95Jy.km/s. 
Braine et al. (1993) with the IRAM-30m measured I(CO)=10 K.km/s on average
over a region of 36\arcsec diameter,
yielding a flux of $\sim$ 120 Jy.km/s. 
In the same region, we measure an average of 
11 K.km/s (and $\sim$ 130 Jy.km/s), which is compatible.
Here with the PdBI
we measure  63 Jy.km/s in the FOV= 43\arcsec\, CO(1-0) map.
We conclude that we have recovered about half of the flux
in CO(1-0). In CO(2-1), with the single dish
we measure 8 K.km/s, compatible with Braine et al. (1993) 
who measured 9 K.km/s, i.e. 
about  240 Jy.km/s in integrated flux, in the inner 26", and the
interferometer measures only 24 Jy.km/s. In this case, about 90\% of
the flux is missing, since the emission is 
not only extended, but also near the edge of our 21" FOV
(the mean radius of the CO ring being 12.5\arcsec\,).

\subsection{IRAM interferometer CO observations}

We observed the emission of the J=1-0 and
J=2-1 lines of $^{12}$CO in NGC\,7217 using the
IRAM Plateau de Bure interferometer (PdBI) in January 2001 (C and D configurations)
and January 2003 (B configuration). 
The six 15\,m antennae were equipped with
dual-band SIS receivers yielding SSB receiver temperatures around
40\,K and 50\,K at the two observed frequencies. The system temperatures were
300\,K for CO(1-0) and 400\,K for CO(2-1). The spectral correlators were centered at
114.906\,GHz and 229.808\,GHz respectively 
(i.e., the transition rest frequency corrected for the galaxy's assumed
redshift derived from V$_{LSR} =$ 952 km/s),
with three correlator units covering a total bandwidth of 400\,MHz
at each frequency. The difference between LSR and heliocentric velocities is 12 km/s, 
and therefore the observations were
centered on V$_{hel}$ = 940 km/s. The
units (each, 160\,MHz wide) provided a nominal frequency resolution of
1.25\,MHz ( $3.25\,{\rm km\,s^{-1}}$ and $1.62\,{\rm km\,s^{-1}}$ for the CO(1-0) and
 CO(2-1) lines, respectively). 
The correlator was regularly calibrated by a noise source inserted in
the IF system.

Visibilities were obtained with twenty one-minute integrations on the source
framed by short ($\sim 2$\,min) phase and amplitude
calibrations on the nearby quasars 2201+315 and 2234+282.
The data were phase calibrated in the antenna-based mode. On
average, the residual atmospheric phase jitter was less than
$30^\circ$ at both frequencies, 
consistent with a seeing disk of 0.6\arcsec--0.8\arcsec\ size and
with a $\sim 5$\% loss of efficiency. The fluxes of the primary
calibrators were determined from IRAM measurements and taken as an input
to derive the absolute flux density scales for our visibilities, 
estimated to be accurate to 10\%.
The bandpass calibration was carried out using 3C273 
and is accurate to better than 5\%.

The data reduction was performed using
the GILDAS software. 
Data cubes with 512$ \times $512 spatial pixels
(0.25"/pixel) were created with  
velocity planes distant by 5 km/s. 
The cubes were cleaned with the Clark (1980) method and restored by 
a 2.4\arcsec$\times$ 1.9\arcsec\ Gaussian beam (with
PA=48$^{\circ}$) at 115GHz and 1.2\arcsec$\times$0.8\arcsec\ (with
PA=54$^{\circ}$) at 230GHz.
The rms noise levels in the cleaned maps (at 5 km/s velocity resolution) are
3\,mJy/beam, and 6\,mJy/beam for the 
CO(1-0) and CO(2-1) lines respectively. No continuum emission was detected toward NGC7217,
 down to rms noise levels of 0.2 mJy/beam and 0.45 mJy/beam in a 580\,MHz bandwidth
at 115GHz and 230 GHz respectively. The conversion 
factors between specific intensity and brightness temperature are
16.7\,K/(Jy/beam) at 115GHz, and 15.6\,K/(Jy/beam) at 230GHz.
 The maps are corrected for primary beam attenuation.

\subsection{Images at other wavelengths}

We acquired from the HST archive four broadband images of NGC\,7217, including
WFPC2 (PI: Smartt; PID=9042) F606W ($\sim$\,V), F814W ($\sim\,$I), and
NICMOS (PI: Stiavelli; PID=7331) F110W ($\sim$\,J), F160W ($\sim$\,H) data.
The optical images were combined with elimination of cosmic rays ({\it crreject}),
and calibrated according to Holtzman et al. (1995).
Sky values were assumed to be zero since the galaxy filled the WFPC2 frame,
an assumption which at most makes an error of $\sim$\,0.1 mag at the corners of the images.
The NICMOS images were re-reduced using the best calibration files
and were subjected to the van der Marel algorithm to remove the
``pedestal'' effect (see B\"oker et al. 1999).
Hereafter these F160W images will be 
referred to as H.
All images were rotated to canonical orientation (North up, East left) using the
rotation angles provided in the headers.


Optical and NIR surface-brightness cuts of NGC\,7217
were derived along the major (PA\,=\,95$^\circ$) and minor (PA\,=\,5$^\circ$) 
axes with a 1\arcsec\ width,
and combined for colors after rebinning onto a common 0.1\,$\arcsec$ pixel scale.
Optical-NIR ($V - H$, $I - H$) color images were obtained by subtracting the magnitude images,
 after rebinning the higher-resolution
NICMOS images to the 0.1\arcsec\ WFPC2 resolution, and aligning them on the galaxy nucleus.
Colors were corrected for Galactic extinction using $A_B\,=\,0.38$ (Schlegel et al. 1998), 
and the extinction curve of Cardelli et al. (1989).
These data will be discussed in Sect. \ref{starform}.

\section{Molecular gas results \label{COres}}

Figure \ref{chan} displays 64 of the channel maps
in the CO(1-0) line, with a velocity range of
320 km/s, and a velocity resolution of 5 km/s. 
The velocity field is strikingly regular,  showing
a classical spider diagram  in the iso-velocity contours
(see also Fig.~\ref{vel}).

\begin{figure}[ht]
\rotatebox{-90}{\includegraphics[width=8cm]{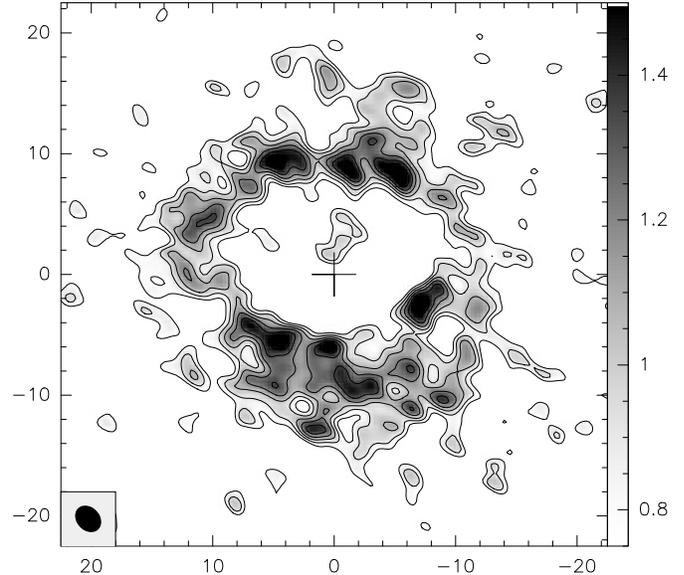}}
\caption{CO(1-0) integrated intensity contours
observed with the IRAM interferometer toward the center of NGC\,7217
(the cross marks the coordinates of the center as given in Table 1,
with offsets in arcseconds).
The map has not been corrected for primary beam attenuation.
The rms noise level is $\sigma$=0.15Jy/beam.km/s.
Contour levels are from 5$\sigma$ to  10$\sigma$ with 1$\sigma$ spacing.
The beam of 2.4" x 1.9" is plotted at the
bottom left.  \label{int}}
\end{figure}

\subsection{Morphology of the CO ring \label{morph}}

\begin{figure}[ht]
\rotatebox{-90}{\includegraphics[width=5.5cm]{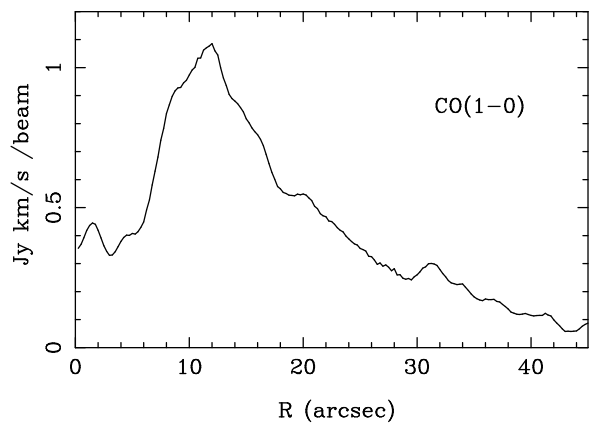}}
\caption{Radial distribution (azimuthal average,
deprojected to face-on orientation)
of the CO(1-0) integrated intensity, shown in 
Fig.~\ref{int}. This shows the abrupt 
drop in intensity at the inner edge of the ring.
\label{radial}}
\end{figure}

The CO(1-0) integrated intensity distribution, displayed in Fig.~\ref{int},
reveals a relatively regular and complete ring, with a 
mean diameter of 25". The depleted region in the center is striking,
 as is the steep gradient of emission at the transition of the
inside of the nuclear ring (see the radial profile
in Fig.~\ref{radial}).
The decrease of the emission is much smoother toward the outer edge of the ring.
Although complete, the ring
consists of individual giant molecular cloud 
complexes, each with a few 10$^6$ to 10$^7$
M$_\odot$, like pearls on a string.
The ring is also asymmetric, appearing wider to the South.

\begin{figure}[ht]
\rotatebox{-90}{\includegraphics[width=8cm]{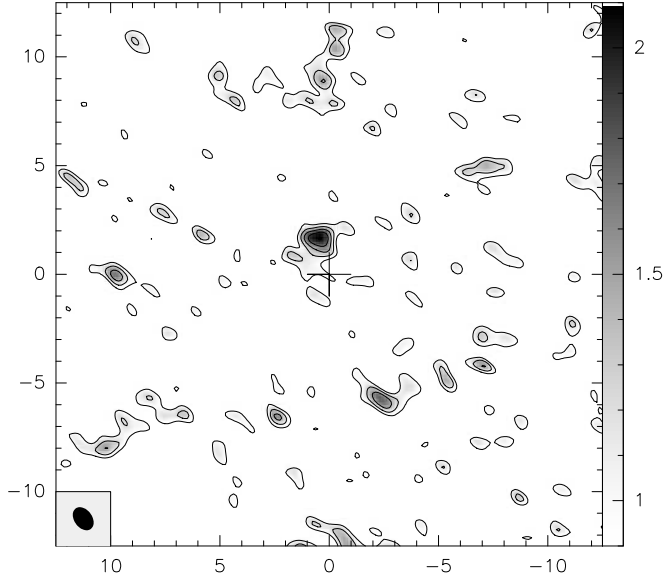}}
\caption{CO(2-1) integrated intensity map
observed with the PdBI toward the center of NGC\,7217
(the cross marks the coordinates of the center as given in Table 1,
with offsets in arcseconds). The central CO(2-1) emission clump coincides with the
  actual galaxy center.
The map has not been corrected for primary beam attenuation.
The noise level is 0.3Jy/beam.km/s.
Contours are from 3$\sigma$  to 10$\sigma$, with 1$\sigma$ spacing.
The beam of 1.2" x 0.8" is plotted at the
bottom left.
\label{int21}}
\end{figure}

Contrary to CO(1-0), the CO(2-1) integrated emission  
is detected mainly inside the ring,
as shown in Fig.~\ref{int21}. The maximum of the emission is in the center,
but the nuclear ring itself is quite weak, and certainly suffers
from the primary-beam tapering off at a HPBW of 22\arcsec, as the CO
ring is 25\arcsec\ in diameter.  When the data are tapered in order to lower
the resolution and increase the brightness sensitivity,
the CO(2-1) ring is better delineated, as shown in Fig.~\ref{21sup10}.

The mass of molecular gas in the nuclear ring from the CO(1-0) 
integrated emission is estimated to be: 

$$ M_{H2} = 3.2\,\times\,10^8 M_\odot$$

\noindent with the standard conversion ratio of
N(H$_2$)/I(CO) = 2.2 10$^{20}$ cm$^{-2}$ (K.km/s)$^{-1}$
(Solomon \& Barrett 1991).

The detection of CO(2-1) inside the ring, with a corresponding
weak CO(1-0) emission at the center (compatible
with a CO(2-1)/CO(1-0) line ratio of 1), implies a mass of
$M_{H2}=4.5\,\times\,10^5 M_\odot$
 inside r $<$ 100pc (the central beam); this is  
not a massive molecular disk around the nucleus, but an amount of gas
equivalent to a single Giant Molecular Cloud. 
The central CO(2-1) (and weak CO(1-0)) peak agree with the
geometrical center of the CO ring, and with the dynamical center of the CO ring.

\subsection{Kinematics of the CO ring\label{kinematics}}

\begin{figure}[ht]
\rotatebox{-90}{\includegraphics[width=8cm]{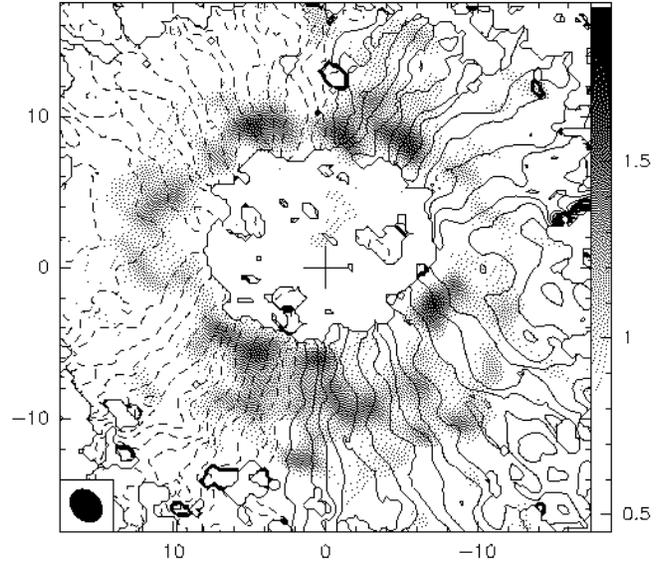}}
\caption{ Overlay of the integrated intensity map of
CO(1-0) (in gray scale) with the CO
mean-velocity field, in contours spanning the range -150km/s to
150km/s in steps of 10km/s.  Velocities are referred to
V$_{hel}$=952 km/s. Solid (dashed) lines are used
for  positive (negative) velocities.  The first solid
contour represents systemic velocity. 
\label{vel}}
\end{figure}

Remarkably, there are no streaming motions, or irregularities
in the velocity field, in accordance with the very
regular and circular nuclear ring.
Isovelocity curves of the CO(1-0) emission are
superposed on the CO emission in Fig.~\ref{vel}. 

\begin{figure}[ht]
\rotatebox{-90}{\includegraphics[height=8cm]{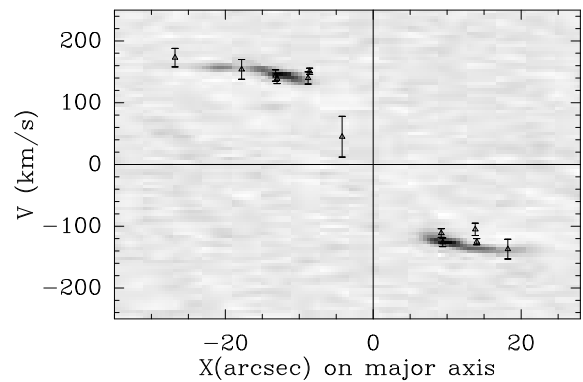}}
\caption{ CO(1-0) position-velocity (p-v)
diagram  along the major axis of NGC\,7217 (East is to the right).
The center is that of Table 1, and the velocity is 
 relative to V= 952km/s LSR (or 940 km/s hel).
The markers with error bars show the H$\alpha$  radial velocities
observed by {\bf B95}.
\label{pv}}
\end{figure}

The position-velocity diagram along the major axis
of the galaxy is presented in Fig.~\ref{pv}. This shows
the regularity of the kinematics, and the sharp decline
of emission inside the nuclear ring.
This confirms the conclusion of B95 that
the {\it flocculent spiral structure}
seen in optical images (see Fig.~\ref{vi-rings}) is {\it trailing},
since the north side is determined to be the near side, 
given the asymmetry in dust obscuration
(see Fig.~\ref{vi-rings}).

\subsection{The excitation of the gas \label{gasexcite}}

\begin{figure}[ht]
\rotatebox{-90}{\includegraphics[width=8cm]{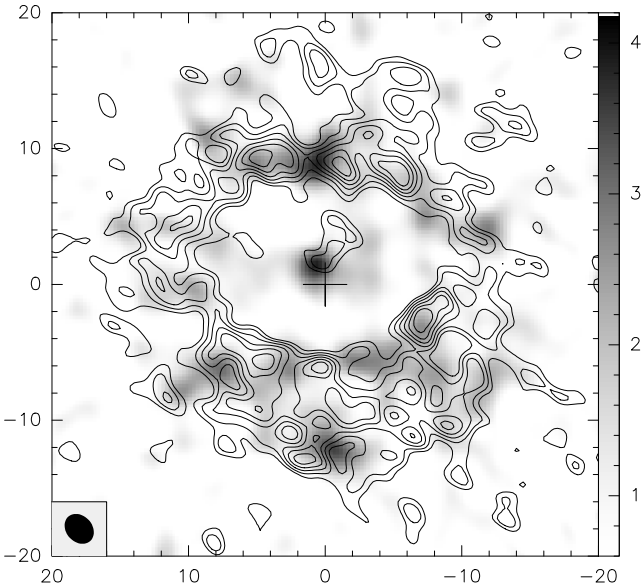}}
\caption{ Contours of the  CO(1-0) map (same as Fig.~\ref{int}), superposed with the
greyscale CO(2-1) map  tapered and convolved to
  identical resolution and corrected for primary beam attenuation, in Jy/beam.km/s. }
\label{21sup10}
\end{figure}

Comparison of the two CO line maps,  at
the same resolution, and with the same 
spatial frequency sampling, can yield information
about the excitation condition of the gas. We have apodised
the CO(2-1) data to sample the same regions of the
 Fourier plane and give the same spatial resolution
at both frequencies.
The result is shown in Fig.~\ref{21sup10}.

\begin{figure}[ht]
\rotatebox{-90}{\includegraphics[width=8cm]{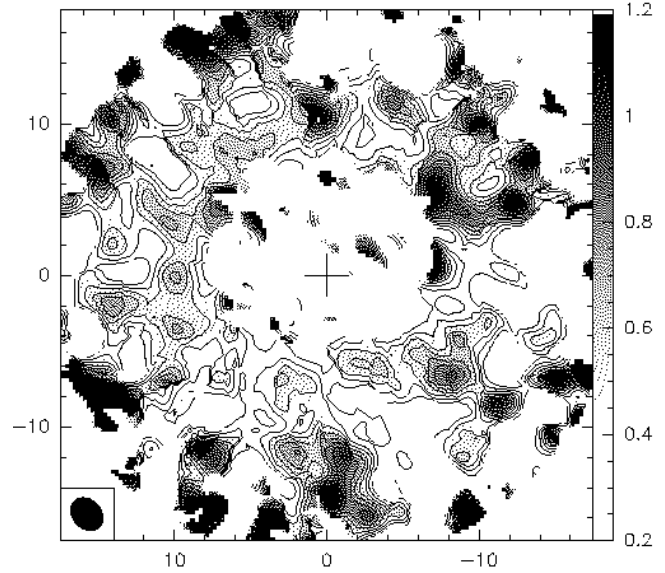}}
\caption{ Contours and grey-scale of the 
CO(2-1)/CO(1-0)ratio map. The ratio is estimated
only when the signals are larger than 3$\sigma$. Contours are
from 0.2 to 1.2 by 0.1.}
\label{21-10ratio}
\end{figure}

The CO(2-1)/CO(1-0) ratio is
consistent with unity in most of the clumps, although with some scatter
(see Fig.~\ref{21-10ratio}). Even the maximum
of 1.4 (in one of the clumps in the ring) is certainly compatible with 
optically thick emission, given the uncertainties. The region
of the molecular ring is indeed very noisy in the CO(2-1) 
map after correction for primary beam attenuation,
since the ring is in the
region where the noise is a factor of two larger than that in the center.
This elevated noise might also 
affect the positions of the CO(2-1) clumps  relative
to those of the CO(1-0) ones.


\subsection{The central CO peak \label{co21}}

\begin{figure}[ht]
\rotatebox{-00}{\includegraphics[width=8.0cm]{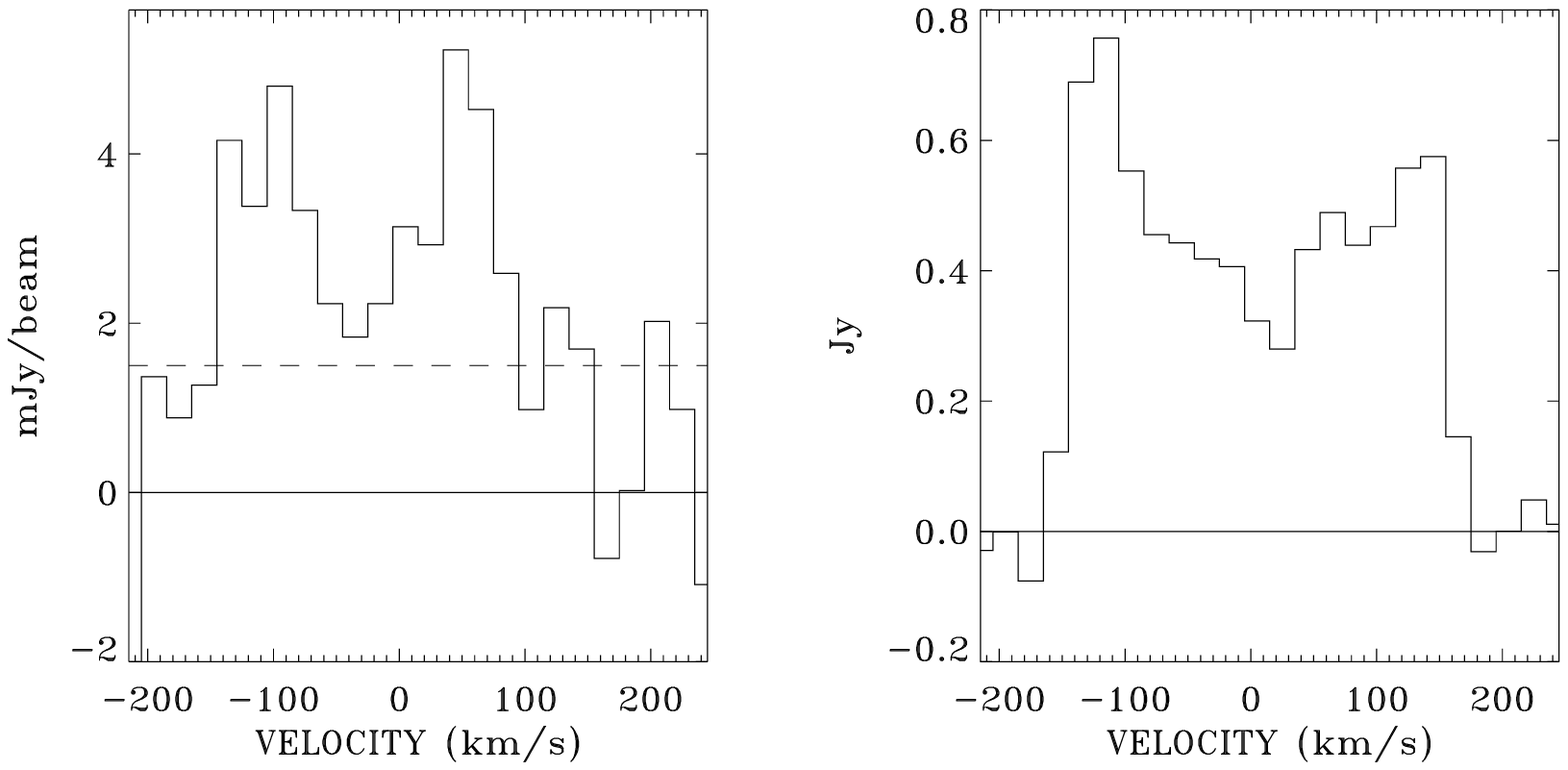}} 
\rotatebox{-00}{\includegraphics[width=8.0cm]{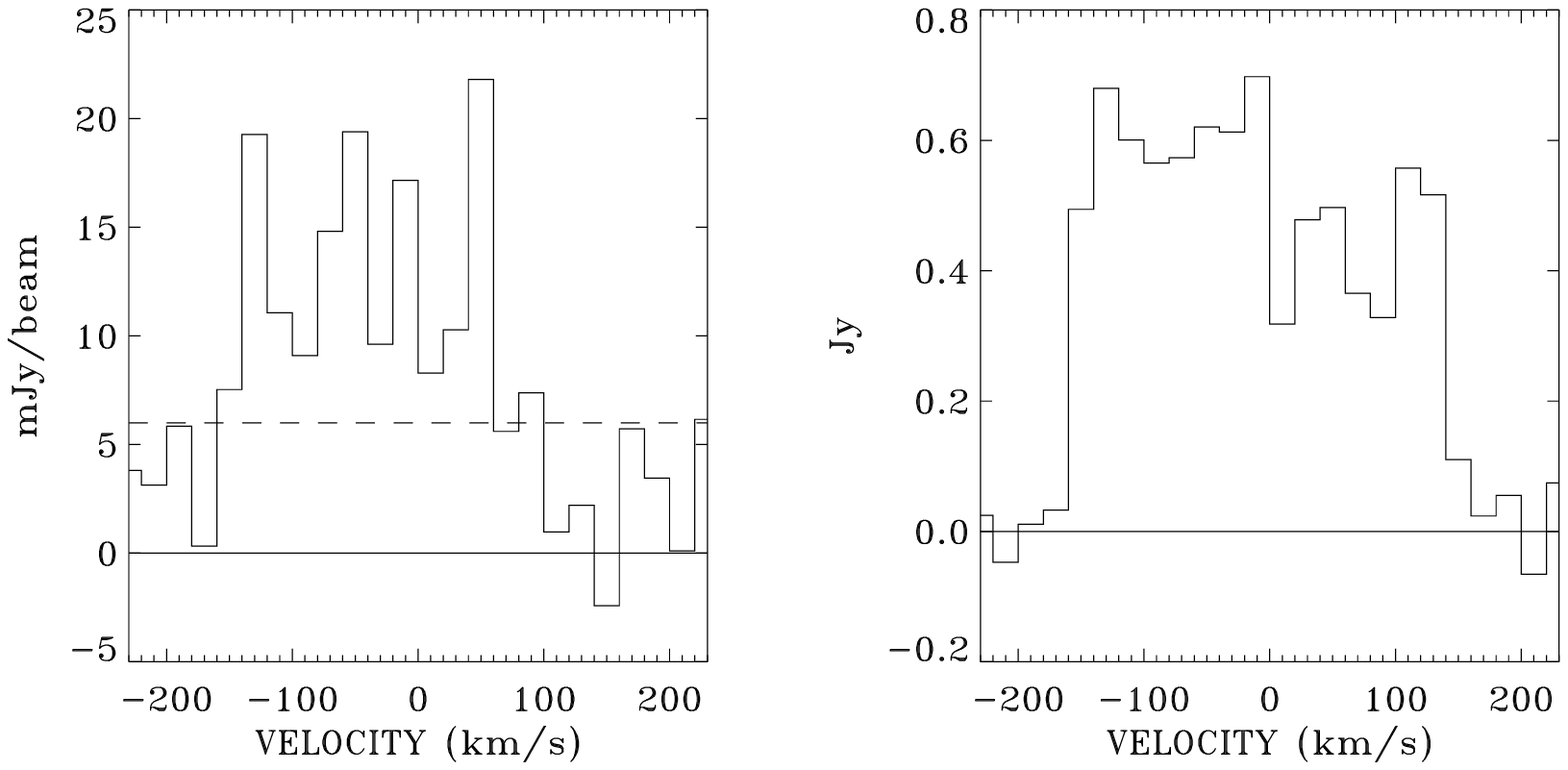}}
\caption{ {\bf Left}
Average CO spectra in the region of maximum emission in the center,
covering a 2.4\arcsec\, x 1.9\arcsec\, beam.
{\bf top:} CO(1-0), and {\bf bottom:} CO(2-1).
The dashed line indicates the 1$\sigma$ level.
{\bf Right}
The spectra summed over the entire CO ring (excluding the center),
{\bf top:} CO(1-0), and {\bf bottom:} CO(2-1).
}
\label{spec}
\end{figure}

The CO(2-1) emission peaks toward the center.
The spectrum in this region is
shown in Fig.~\ref{spec}. The emission is spatially unresolved
and corresponds to the
tentative detection of a compact disk inside r=2\arcsec.
There is also a weak CO(1-0) counterpart.
As mentioned in Section 3.1 however, the molecular mass included in this
central spot is only that of one GMC (4.5 \, 10$^5$ M$_\odot$). 


There may be a link between the circumnuclear ionized gas and our 
detection of a compact CO(2-1) source.
Sil'chenko \& Afanasiev (2000) claimed to detect a very central polar ring 
structure in ionized gas roughly aligned in North-South direction. 
Their claim of a polar geometry rests on
the twisted isovelocity curves in the center.
However, this kind of S-shape in the isovelocity curves is rather 
common in galactic nuclei, and is generally
 due to non-circular orbits along a bar
or spiral arms. Only weak CO(1-0) 
emission is detected inside the nuclear ring; the present
molecular observations do not provide evidence of abundant AGN fueling at the 100pc scale 
in NGC7217 at present. Some estimate of the accretion luminosity (from the H$\alpha$
 line flux, cf Ho et al. 1997) indicates that fueling was effective
in a recent past. 

\begin{table}[ht]
\caption{ Different rings in NGC\,7217 \label{ringtab} }
\begin{flushleft}
\begin{tabular}{lll}
\hline
Parameter  &   Value                              &  Reference \\
\hline
Radius of outer ring &  77\arcsec\ =  5.4 kpc          & B95 \\
Radius of inner ring &  32\arcsec\ =  2.2 kpc          & B95 \\
Radius of red ring &  14\arcsec\ =  0.98 kpc          & V95 \\
Mean radius of CO ring& 12.5\arcsec\  = 0.9 kpc      & this work \\
Stellar nuclear ring & 10.7\arcsec\ =  0.75 kpc   & B95 \\
H$\alpha$ nuclear ring radius& 10.7\arcsec\ = 0.75 kpc  & P89 \\
Dust ring ($V - I$)       &  8.6\arcsec\ =  0.6 kpc          & B95 \\
\hline
\end{tabular}
\end{flushleft}
B95: Buta et al. (1995)\\
V95: Verdes-Montenegro et al. (1995) \\
P89: Pogge 1989, and also Verdes-Montenegro et al. (1995) \\
\end{table}

\section{Comparison to other wavelengths \label{starform}}

\begin{figure}
{\includegraphics[width=8.5cm]{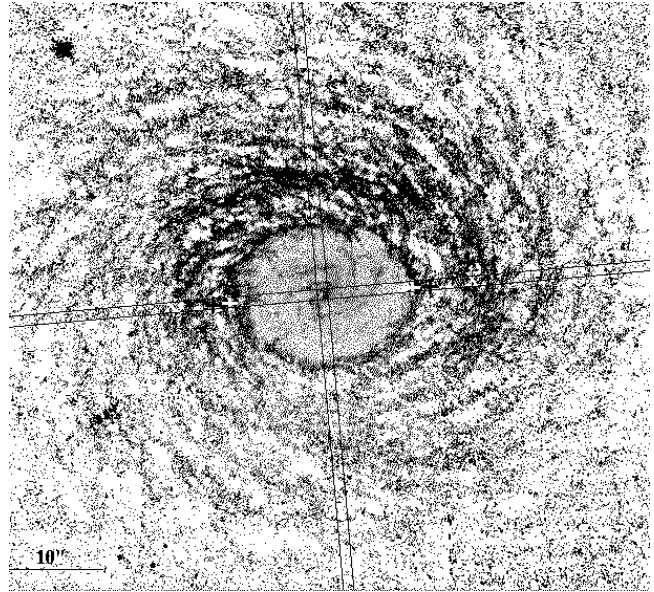}}
\caption{HST $V - I$ (transformed from F606W-F814W) image with major- and
minor-axis cuts superimposed.
The positions of the rings are shown as a
{\it i)}~white cross = red dust CNR;  
{\it ii)}~black cross = blue stellar CNR, H$\alpha$ ring;  
{\it iii)}~white $\times$ = pseudo CNR;  
{\it iv)}~black $\times$ = blue stellar inner ring.  
Darker pixels correspond to redder colors, and 
lighter pixels to bluer. The total horizontal size of this
image is 67'' (North is up and East is left). Same color scale 
as in Fig.~\ref{CO10-vi}. 
\label{vi-rings}}
\end{figure}

It is of interest to compare the position, size, and symmetry of the CO
nuclear ring with those delineated by tracers at other wavelengths
(cf. Table \ref{ringtab}).
Previous ground-based observations revealed, in addition to a blue outer ring
at a radius of $\sim\,$77\arcsec\ and a blue inner ring of
radius $\sim\,$32\arcsec, a series of {\it circumnuclear
rings} (CNRs):
\begin{enumerate}
\item ~a red ring at radius $\sim\,$14\arcsec\ (denoted as ``inner red ring''
by Verdes-Montenegro et al. 1995);
\item ~a mildly blue (in $V - I$) stellar ring at radius $\sim\,$11\arcsec\ (B95,
Verdes-Montenegro et al. 1995); 
\item ~an H$\alpha$ ring (Pogge 1989) coincident with the blue stellar ring; and
\item ~a red $V - I$ dust ring at a radius of $\sim\,$8.6\arcsec (denoted
as ``nuclear dust ring'' by B95).
\end{enumerate}
The ``inner red ring'', at radius $\sim\,$14\arcsec, is relatively broad (210pc, $\sim\,$3\arcsec) 
and incomplete, is defined by flocculent spiral structure, and appears
as a semi-circle with no red emission to the South; it will be
referred to hereafter as the ``pseudo CNR''.
In contrast, the better defined narrower ($\leq$\,70pc, 1\arcsec) circumnuclear red 
dust ring at radius $\sim\,$8.6\arcsec\ forms a
closed oval structure, as do the slightly larger stellar and H$\alpha$ rings.
Figure \ref{vi-rings} shows the positions of these rings in cuts
superimposed on the HST $V - I$ image; the cuts will be described in Section 4.3.
Our HST $V - I$ image clearly resolves all the rings, and reveals additional 
substructure (e.g., blue star
 clusters) which was previously not obvious from ground-based images.

\subsection{Circumnuclear rings}

\begin{figure}
\rotatebox{-90}{\includegraphics[width=7.5cm]{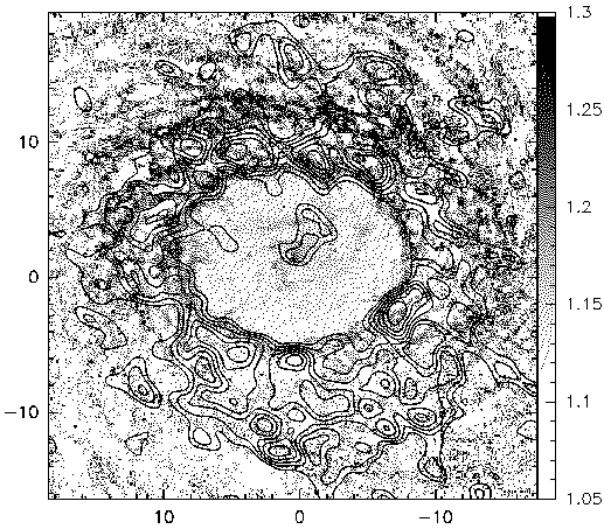}}
\caption{ CO(1-0) low-resolution contours superposed on the HST $V - I$ color image. 
 Darker pixels correspond to redder colors, and
lighter pixels to bluer.
\label{CO10-vi}}
\end{figure}

Figure \ref{CO10-vi} shows the superposition of the CO(1-0) contours onto
the HST $V - I$ map. 
$V - I$ essentially traces extinction but blue colors of young stars
can also be seen just outside the red dust CNR. 
The molecular ring starts abruptly at exactly the position of the inner
well-defined dusty CNR (radius 8.6\arcsec, see also Fig. \ref{radial}),
and covers the stellar and ionized gas CNRs, out to
the flocculent trailing spiral structure in the pseudo CNR. 
The bulk of the CO(1-0) molecular ring is coincident with, although broader than, 
the CNRs revealed by other tracers.
Indeed the CO(1-0) integrated intensity map shows a 
ring significantly broader than either the dust or stellar CNRs.
If fitted by a Gaussian, the half-power width is of order
570pc ($\sim\,$8\arcsec); measuring the half-power width directly
gives an even larger value, closer to 700pc ($\sim\,$10\arcsec) (see Fig. \ref{radial}). 
The CO resolution is not sufficient to
separate the nuclear ring itself from the pseudo CNR and
flocculent spiral structure that blends with it at larger radii.
The sharp cutoff of the CO ring is conspicuous 
on its inner edge, but then the ring merges with
the spiral structure outside, visible in the dusty
flocculent features, and well resolved by the HST color image. 

\subsection{Nuclear spiral and blue nucleus }

\begin{figure}
\rotatebox{-90}{\includegraphics[width=7.5cm]{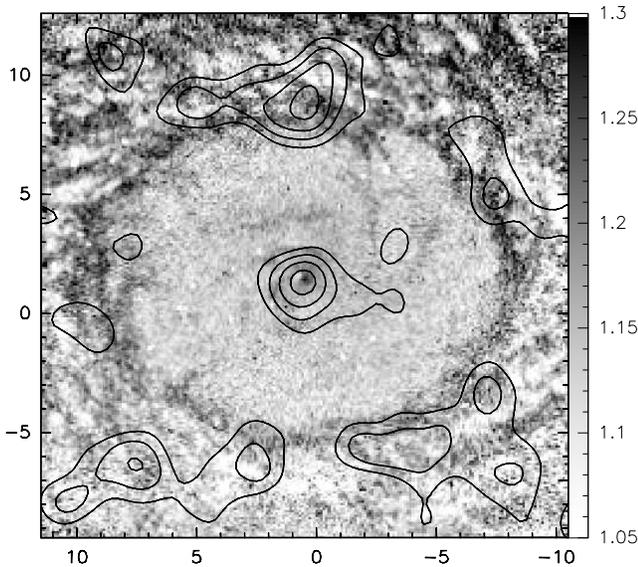}}
\caption{ CO(2-1) contours superposed on the HST $V - I$ color
image. As in Fig. \ref{CO10-vi}, darker pixels mean redder colors.
The CO(2-1) peaks
exactly at the center of the nuclear bar/spiral feature.
\label{CO21-vi}}
\end{figure}

At the center of the galaxy and its CNR,
there is a small multi-arm spiral  dust lane encircled by the larger flocculent one.
Figure ~\ref{CO21-vi} shows the superposition of the CO(2-1) contours on
the HST $V - I$ map. 
The CO(2-1) contours in the very center coincide with the 
dusty mini-spiral delineated by the red colors in the HST $V - I$.

\begin{figure}
\rotatebox{-90}
{\includegraphics[width=7.5cm]{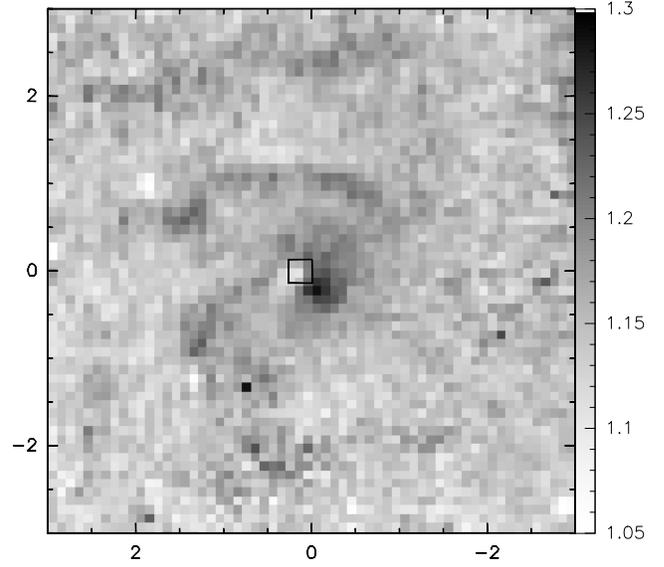}}
\caption{Zoomed $V - I$ image of center of NGC\,7217 with continuum
brightness peak (in both F606W and F814W) indicated with an open square.
Darker pixels correspond to redder colors, and lighter to bluer ones.
Axes are in arcsec. The center (0,0) is now the galaxy center.}
\label{ctr}
\end{figure}

The nucleus turns out to be rather blue in $V - I$ ($V - I$\,=\,1.2),  $\simgt\,$0.3\,mag
{\it bluer} than a red knot ($V - I$\,=\,1.5) to the 
southwest. Figure  \ref{ctr} shows a close-up of the center of NGC\,7217 in
$V - I$; the position of the continuum peak
is shown as an open square.
This red knot appears to be the center of the nuclear spiral, which
also has red colors (although 0.1--0.15\,mag bluer than those of the knot)
and appears to be dusty.
The blue nucleus and the red knot are separated by roughly 0.3\arcsec,
or $\approx\,$24 pc.
The blue colors of the nucleus subtend a region of 28\,pc$\times$20\,pc,
with a position angle of $\sim\,30^\circ$.

 The presence of gas and dust in a spiral structure inside the 
nuclear ring suggests that some gas is still able to make
its way towards the nucleus through viscosity, in spite of the 
contrary action of the gravity torques.

\subsection{North-south asymmetry}

As mentioned previously,
the broad CO ring is roughly coincident with the narrow stellar and H$\alpha$ ring 
at $\sim\,$11\arcsec.
The CO(1-0) emission starts just on the red dust ring at a radius of 8.6\arcsec;
and is very broad, almost ten times wider than the red dust or stellar rings.
Interestingly,
while the stellar and red dust rings are closed structures,
the pseudo CNR just outside the stellar ring appears as a semi-circle, open
in the direction where CO is broader.
The implication is that the asymmetry of the red pseudo CNR
cannot be ascribed to an intrinsic hole in the dust, 
since the CO emission is on the contrary even more conspicuous toward the south. 

\begin{figure}
{\includegraphics[width=9cm]{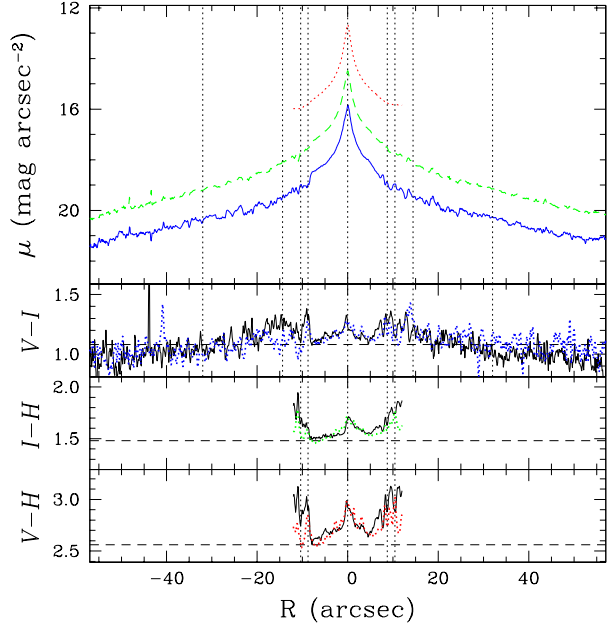}}
\caption{Surface brightness cuts taken along the slits shown in Fig. \ref{vi-rings}.
The dotted line corresponds to F160W, the dashed line to I, and the solid 
line to V.
In all panels, the various ring positions are shown by vertical dotted lines
(from center: red dust ring, stellar ring, pseudo CNR, and inner stellar ring).
In the lower panels, major-axis cuts are shown as solid lines and minor-axis cuts
as dotted lines;
horizontal dashed lines illustrate the colors of a 26\,Myr stellar population
reddened by 1.1\,mag of extinction (see Section \ref{starform}).
The minor-axis cuts have been rectified by using an axial ratio of 0.8.
}
\label{cut}
\end{figure}

The north-south (N-S) asymmetry can be seen  more clearly in Fig. \ref{cut},
which shows major- and minor-axes cuts and their colors, taken at the positions
shown in Fig. \ref{vi-rings}.
In the figure, dotted curves show the minor-axis cuts rectified to circular shapes
(inclination = 36$^\circ$; PA= 95$^\circ$).
Also shown as vertical dotted lines are the positions of the various rings. 
Inspection of the figure shows clearly the effect of the flocculence on the colors
(ripples in the color curves)
and that the $V - I$ color (see short-dashed profile)
is significantly bluer toward the S (negative galactocentric
distance) than toward the N at the position of the pseudo CNR at $\sim\,$14\arcsec.
Paradoxically, it is toward the South, suffering less extinction, that the
CO emission is stronger, and the molecular ring more massive.

The reasons for such a north/south asymmetry are not clear, but
perhaps the flocculent spiral structure
widens the CO nuclear ring, which is 
more conspicuously stochastic in the south.

\section{Circumnuclear star formation \label{starform}}

\subsection{Colors, ages and extinction }

To better characterize the star formation history and the ages of the stellar 
populations in NGC\,7217, we have measured
the $V - I$, $I - H$, and $V - H$ colors of several circumnuclear regions:
the ``red knot'' in the nuclear spiral,
the (blue) nucleus, the region inside the dust CNR,
the dust CNR itself, and the blue stellar ring.

We have performed a least-square fit of the colors for each region to
the solar metallicity evolutionary synthesis models by Leitherer et al. (1999, hereafter SB99,
dealing with a single burst and Salpeter IMF).

\begin{figure}
\rotatebox{-90}{\includegraphics[width=8cm]{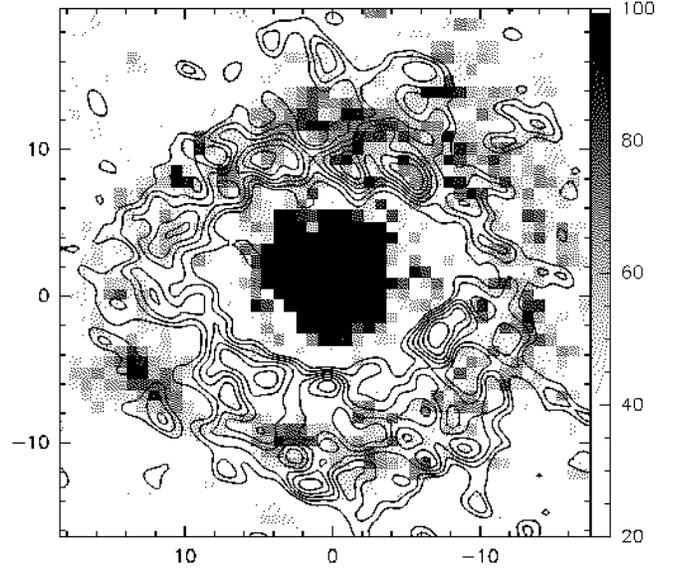}}
\caption{ Contours of CO(1-0) emission 
 (same as Fig.~\ref{int}), superposed on the continuum subtracted
H$\alpha$+[NII] gray-scale map from Pogge (1989). 
Offsets are in arcseconds.}
\label{Halpha}
\end{figure}

We conclude that the blue nuclear stellar ring and the regions encircled by it
are dominated by relatively young stellar populations (10--50\,Myr), with varying amounts of
extinction ($A_V\,\sim\,1-2$\,mag).
These are not typical features of early-type spiral galaxies, and are probably the 
result of a significant gas reservoir and the ensuing massive star formation. 
Indeed, ionized gas emission,
traced by the gray-scale H$\alpha$ map underneath the CO(1-0) 
contours in Fig.~\ref{Halpha}, shows that 
star formation is still occurring in the nuclear regions (nucleus, red knot,
multi-armed spiral), and in the CO ring (although there is
in the center the contribution of the AGN).
The ionized gas traces very well the blue stellar ring, but is absent in the red
dust CNR, probably because of the large extinction there.
However, observations of  recombination lines at longer wavelengths are needed
to better quantify the dust distribution, and hence to confirm
our previous conclusion.

\subsection{The nature of the nuclear spiral }

The nuclear multi-armed spiral in NGC\,7217 revealed by the HST images is similar 
to those in other early-type flocculent galaxies (e.g. Martini et al. 2003).
It is not an obvious continuation of the structure outside the CNR, which
if it corresponds to an inner Lindblad resonance (ILR),  should
shield the nuclear region from incoming spiral waves (Bertin et al. 1989).
However, it is affected by \simgt 1\,mag of dust extinction, and its colors
suggest that it hosts a recent (20-40\,Myr old) starburst.
In other flocculent galaxies (Elmegreen et al. 1998, 2002), 
such small-scale spiral structure has been attributed to acoustic
turbulence and implicated in the fueling of the LINER activity.
 Such spiral pieces could only be material arms, e.g. giant molecular clouds, sheared
off by differential rotation.

Following Thornley (1996), we have measured the ratio of the nuclear spiral arms in the I and
H bands relative to a symmetric model.
The nuclear spiral in the NICMOS F160W appears in an unsharp-masking image, although
with very weak amplitudes, ranging from 1.01 to 1.05.
The I band shows similarly small excursions.
Since near-infrared amplitudes of typical density-wave spirals range from 1.5 to $\sim$3.0
(e.g., Rix \& Zaritsky 1995), we conclude that the nuclear spiral in NGC\,7217 cannot
be a stellar density wave. 
 The presence
of gas and dust in this feature may show how gas is brought to the nucleus.

\section{Dynamical model of the galaxy}

\subsection{Overview}

To summarize the essential features found in the previous sections,
the molecular gas in NGC\,7217 has a very
regular morphology and is concentrated in a nuclear ring.
This ring's most striking feature is its extremely
 sharp inner edge.
When the molecular gas distribution is compared to the HST $V - I$ image, it becomes
clear that the broad CO distribution is composed of a narrow ring,
coincident with the stellar and H$\alpha$ nuclear ring of 1.5kpc diameter,
and a broader ``ring'' of flocculent spiral arms extending to
larger radii, with a width of $\sim$ 400pc.  This complexity explains the apparent
radial shift of average ring radii among the various components,
as noted in Table \ref{ringtab}. The CO ring and spiral arms are very clumpy,
with dense gas in giant molecular complexes (with a few 10$^6$ to 10$^7$ M$_\odot$).

There is a mini spiral structure inside the ring, gaseous and dusty, 
which contains only a very small amount of gas (of order
5 10$^5$ M$_\odot$). With this small mass spread over a relatively
 large area (i.e. surface density of the order of 0.3 M$_\odot$/pc$^2$),
it is likely that this structure is not a
 single self-gravitating unit.
At 1kpc radius, the Toomre critical surface density for axisymmetric
instabilities is about 300 M$_\odot$/pc$^2$; inside the ring, however,
the average molecular surface density is 1 M$_\odot$/pc$^2$
(individual clouds can still be self-gravitating).
This gas could flow inward very slowly
 due to viscous torques.

The most likely interpretation of the global structure is in terms of resonant rings
produced by a bar structure that has faded into an oval 
distortion by now (e.g. B95).
Even the slight oval observed today is able to form the observed rings, as shown by
the simulations of gas flow in the potential derived from a red image
(B95). 
The main gravity torques inside 
the nuclear ring are positive, driving the gas out toward the ring boundary.
However, even if there were
a bar some time ago, the gravity torques are now weak, and
some gas could still flow in through viscous torques, 
explaining the presence of the multi-armed spiral inside the ring.

Given the weakness of the oval distortion observed now,
the abrupt cutoff at the inner edge of the nuclear CO ring is difficult 
to explain.
Two solutions can be considered: either a strong bar formed the rings in the
recent past, and the present morphology shows the residual signatures of that
 process, or there 
have been only weak oval distortions recently,
 one of which is still at play today.
We explore these possibilities, with the help of N-body simulations.
This self-gravitating model is the next step in the interpretation 
proposed by B95, who computed
only the gas response in a fixed potential, at a fixed time. 
Our aim here is to find a plausible dynamical scenario to
account for the presently observed gravitational potential; this requires that we follow
the evolution of all components, including the evolution of the gas content,
and therefore star formation and feedback are included. 
In contrast to B95, we will focus on the central region, where we have  many new
observational constraints coming from the high-resolution
HST images and the interferometric CO data.
Only the formation of the nuclear ring is 
studied, and our examination of the gas is confined to the center. We do not follow
the formation of the more external rings, since these would require external gas
accretion, that has been ignored here.

\subsection{Physical model adopted for the galaxy and numerical techniques \label{code}}

In order to understand the observed morphology, we
performed N-body simulations with stars and gas, including star formation.
The regularity and axisymmetry of the bulge
and stellar halo components led us to consider them as spherical,
and rigid potentials, in which the disk component evolves
(e.g. Sellwood 1980).
Self-gravity is only included for the disk (gas + stars).
2D N-body simulations were carried out 
using the FFT algorithm to solve the Poisson equation, with
a polar grid, to  optimize the spatial resolution toward
the center.

The polar grid is composed of $NR=64$ radial, and $NT=96$ azimuthal 
separations. While all azimuthal increments are equal,
the radial grid is exponential, with the cell size ranging from 17
 pc at the center to $\sim 1\,{\rm kpc}$ at the outskirts
(at 15kpc).
The softening is also variable, ranging from 50\,pc near the center to 
1kpc.
The number of cells is chosen so that cells are approximately
square at all radii (almost equal size in radius and azimuth).
Radii of the grid points are given by:
$$R_I = 0.25 \, exp(2 \pi I /NT)$$
\noindent for $I$ =1 to 63; and $R_0=0$ for $I=0$.
The self-force introduced by the polar grid is
subtracted at each time-step, as already described
in Combes et al. (1990).

The stellar component
is represented by 80000 particles, and the gas component by 40000.
When an extended bulge component is added, to control the disk stability
and the bar strength, it is a rigid spherical potential, with a Plummer shape:
$$
\Phi_{EB}(r) = - { {G M_{EB}}\over {\sqrt{r^2 +r_{EB}^2}} }
$$
for $M_{EB}$ and  $r_{EB}$ the mass and characteristic radius of the
extended bulge component respectively. To fit the rotation curve of
NGC\,7217, however, the characteristic radius of this component
has to be small, 3.5 kpc. This component may be considered 
either as the extended bulge or the stellar halo that has been previously noticed 
 in this galaxy (e.g. B95).

The stellar disk is initially a Kuzmin-Toomre disk of surface density
$$
\Sigma(r) = \Sigma_0 ( 1 +r^2/r_d^2 )^{-3/2}
$$
truncated at 14kpc, with a mass  M$_d$.
It is initially quite cold, with a Toomre Q parameter of 1.
The bulge is a Plummer sphere, with mass M$_b$ and characteristic radius
r$_b$.
The time step is 0.1 Myr. The initial conditions  of the runs 
described here are given in Table \ref{condini}.

\begin{table}[ht]
\caption[ ]{Initial conditions parameters}
\begin{flushleft}
\begin{tabular}{ccccccc}  \hline
Run       & r$_b$& M$_b$    & M$_d$    & M$_{EB}$$^*$& F$_{gas}$& $f_{el}$  \\
          & kpc  & M$_\odot$& M$_\odot$& M$_\odot$   &  \%      &        \\
\hline
Run A     &  0.5 &  2.7e10 &  4.5e10   & 6.0e10      &  3       & 0.35    \\
Run B     &  0.7 &  3.6e10 &  7.0e10   &  0.         &  1.5     & 0.65    \\
\hline
Run A1    &  0.5 &  2.7e10 &  4.5e10   & 6.0e10      &  6       & 0.65    \\
Run A2    &  0.5 &  2.7e10 &  4.5e10   & 6.0e10      &  1.5     & 0.35    \\
Run A3    &  0.6 &  3.2e10 &  5.0e10   & 4.5e10      &  3       & 0.65    \\
Run A4    &  0.7 &  3.6e10 &  5.0e10   & 5.0e10      &  3       & 0.35    \\
Run A5    &  0.8 &  3.6e10 &  3.6e10   & 6.0e10      &  3       & 0.65    \\
\hline
Run B1    &  0.5 &  2.7e10 &  6.0e10   &  4.5e10     &  6.      & 0.65    \\
Run B2    &  0.5 &  2.7e10 &  6.0e10   &  4.5e10     &  3.      & 0.35    \\
Run B3    &  0.5 &  2.7e10 &  7.7e10   &  2.7e10     &  3.      & 0.35    \\
Run B4    &  0.5 &  2.7e10 &  7.7e10   &  2.7e10     &  6.      & 0.65    \\
Run B5    &  0.5 &  2.7e10 &  1.0e11   &  0.         &  6.      & 0.85    \\
\hline
\end{tabular}
\end{flushleft}
$^*$ Extended Bulge (EB) mass inside 14 kpc radius\\
r$_d$ and  r$_{EB}$ are fixed at 3.5kpc\\
\label{condini}
\end{table}

The gas is treated as a self-gravitating component in the N-body
simulation, and its dissipation is treated by a sticky particle code,
as in Combes \& Gerin (1985). The initial gas-to-total mass ratio 
(F$_{gas}$)  ranges
between 1 and 6\%,  since the star formation in the simulation is
 capable of reducing $F_{\rm gas}$ to its observed low value.
The mass of one gas particle therefore varied between 4 \,10$^4$ and 2 \,10$^5$ M$_\odot$.
At the present time, the observed gas mass in NGC\,7217
is only 0.8 10$^9$ M$_\odot$, representing about
1\% of the total mass inside 10kpc required to fit the observed rotation
curve. The initial distribution of gas in the model is an exponential
disk, truncated at 12\,kpc, and with a characteristic radial scale of 2\,kpc.
Initially, its velocity dispersion corresponds to a Toomre Q-parameter
of 1. The gas clouds are subject to inelastic collisions, with a collision
cell size of 240pc (region where particles are selected
to possibly collide). This corresponds to a lower limit for the average mean
 free path of clouds between two collisions. The collisions are considered
every 5 to 10 Myr. In a collision, the sign of the relative
cloud velocities is reversed and the absolute values are reduced: 
relative velocities after the collision are only $f_{el}$ times their original value,
the elasticity factor $f_{el}$ being between 0.35 and 0.85, as indicated 
in Table \ref{condini}.
The dissipation rate is controlled by this factor.
All gas particles have the same mass.

Star formation is taken into account following a generalised
Schmidt law: the star formation rate is proportional to the
volume density to the power n=1.2, provided that the density
is larger than 1 H-atom cm$^{-3}$, i.e. the rate of gas mass
transformed into stars is $dm = dt  C_* \rho^{1.2}$.
To compute this rate, at regular intervals of dt= 5-10 Myr, the gas density
is averaged in 240pc cells, and the probability of the gas particles  being
transformed into stars is computed by
$$
P = dm/M_{cell}
$$
for all particles in this cell, of mass $M_{cell}$.  Each new star formed has exactly the
same mass as each gas particle, about 5 times smaller than any old
stellar particle.  This simple scheme corresponds to an instantaneous
recycling of matter, since the continuous mass-loss from recently formed stars
is not followed.  The rate of star formation is normalised
so that in unperturbed runs (without galaxy interaction, galaxies are
quiescently and regularly forming stars), the timescale for consumption of half of the gas mass
is of the order of 2 Gyr (SFR $\sim$ 1-2  M$_\odot$/yr). 
At each star formation event, the neighbouring gas particles are given
a small extra velocity dispersion of order $\sim$ 10 km/s. 

\begin{figure}[ht]
\rotatebox{-90}{\includegraphics[width=6cm]{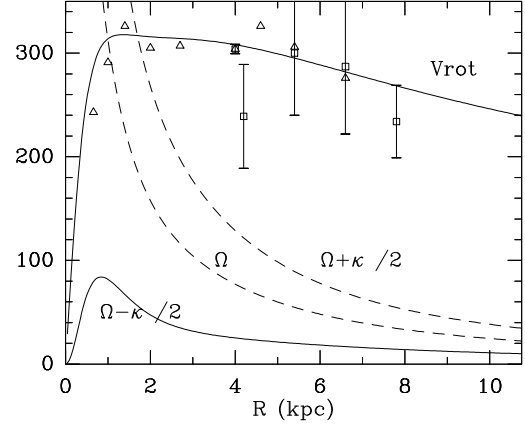}}
\caption{ Rotation curve and derived frequencies $\Omega$,
$\Omega-\kappa/2$ and $\Omega+\kappa/2$ adopted 
for the simulations (Run A). The triangles are the H$\alpha$
rotational velocities; the squares are from the HI data (see B95).}
\label{vcur}
\end{figure}

The rotation curve corresponding to one of the runs
is plotted in comparison to the data points in Fig.~\ref{vcur}.
All other runs have similar rotation curves.
Given the uncertainties, all curves are compatible with the data.

\section{Simulation results}
\label{simul}

 About 50 runs have been carried out, to test the various parameters,
and evaluate the dependence of the results
 on dissipation rate of the gas, and the initial total 
mass of the gas. These are crucial in determining the disk stability,
and the strength of the developed bar or oval distortion.
 Only a dozen of these models are displayed in Table 3.
 
The stellar disk self-gravity is the most essential parameter
affecting the stability and bar
 strength.  To restrict the volume of parameter space to explore, we have
 chosen to fix the scale lengths of the mass components, and mainly vary only
 the {\it masses} of two components (the disk and extended bulge).
Only two representative runs (A and B),
will be discussed here; their mass parameters are displayed in 
Table \ref{condini}. 
These two runs are typical
of two categories of models, A$i$ leading to weak bars or oval distortions,
and B$i$ to strong bars.

Run A, with a maximal bulge mass, provides the best fit for NGC\,7217.
Run B is the maximum disk
solution, where there is no extended bulge component. The strength of 
the $m=2$ perturbation increases steadily as the mass ratio between
disk and bulge is increased. However, for a given mass ratio,
the $m=2$ strength also varies significantly with
the initial gas mass (i.e., $F_{\rm gas}$, the disk being more
 unstable with larger gas mass fraction),
 and the dissipation rate, controlled
by the elasticity of collisions ($f_{el}$).
The more dissipation, the cooler the disk, the more prominent the bar, and
 the greater the departure of the gas behaviour from that of the stellar
 component, with the gas having an increasing tendency to form rings.
With less dissipation, the rings are less contrasted, and
more of the gas particles share the bar morphology.

\bigskip

Given the present unbarred structure of NGC\,7217, there could be two
possibilities to account for the present ring morphology.
One possibility is that the galaxy was strongly barred in the recent past, 
that gaseous rings have developed, and that new stars have formed in 
these rings;  then
the bar has disappeared, but the rings have remained.
This solution requires a large mass of gas, so that the radial gas
flow driven by the bar can indeed build a mass concentration
of at least 5\% of the total mass (Hasan et al. 1993). 
The other possibility is that the rings have developed through
the action of an oval distortion alone, which is still present (B95).
This solution has the advantage
of explaining the abrupt inner edge of the nuclear ring,
since gravity torques would still be present to compensate
for the natural tendency of the gas to diffuse and spread radially 
out of the rings.

\begin{figure*}
\rotatebox{-90}{\includegraphics[width=5cm]{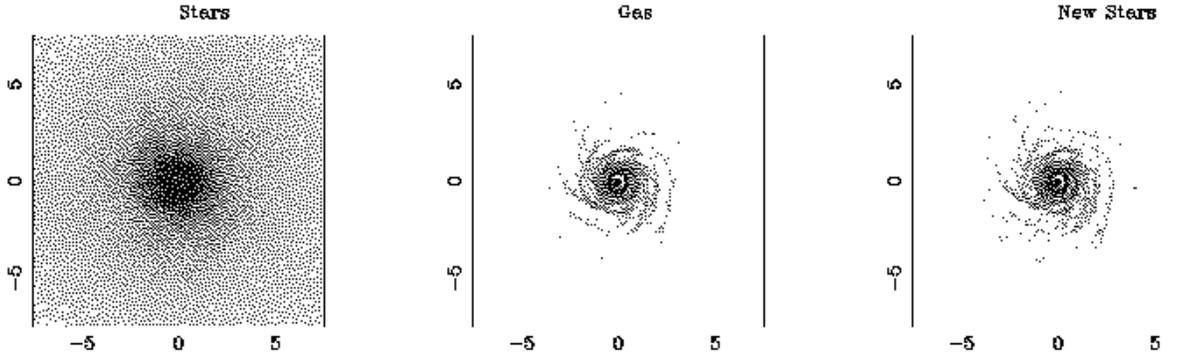}}
\caption{ {\bf Left:} Gray scale plot of the stellar component
distribution, in Run A (the gray scale is linear and axes are in kpc).
{\bf Middle:}  Gray scale plot of the corresponding gaseous component
(the gray scale is logarithmic).
{\bf Right:}  Plot of the stars formed during the simulation
(the gray scale is logarithmic).}
\label{simplot}
\end{figure*}

\begin{figure*}
\rotatebox{-90}{\includegraphics[width=5cm]{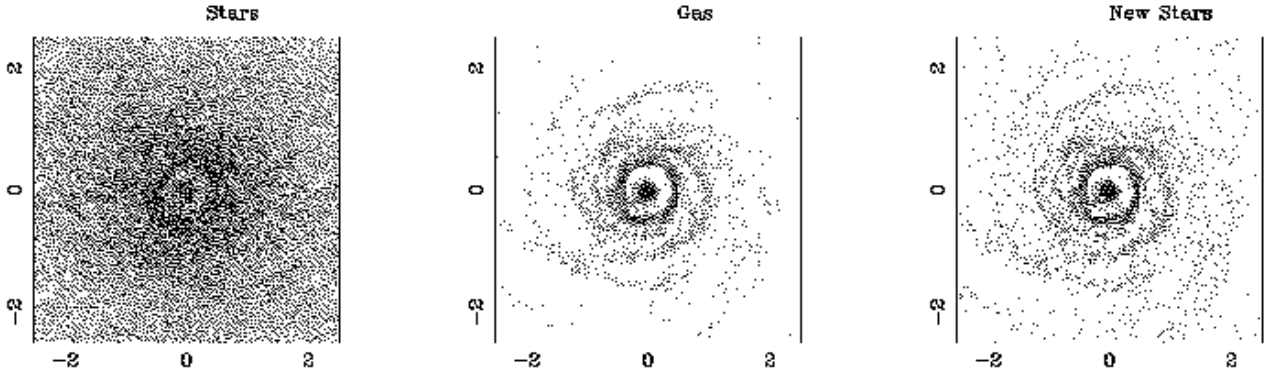}}
\caption{ Same as Fig.~\ref{simplot}, but zooming toward
the center (the axes are in kpc).} 
\label{simplot2}
\end{figure*}

\begin{figure}[ht]
\rotatebox{-90}{\includegraphics[width=6cm]{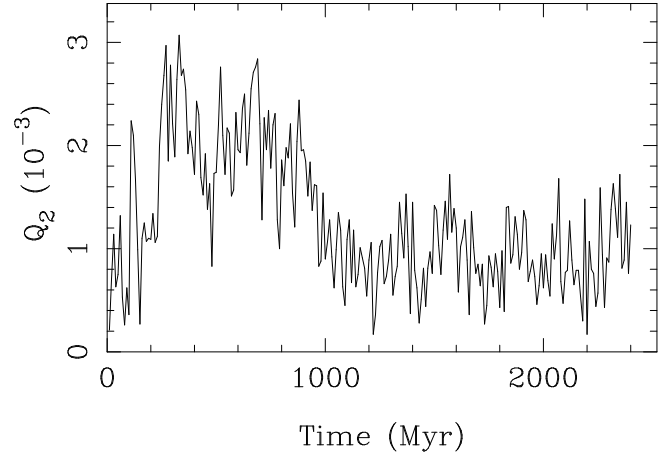}}
\caption{ Evolution of the bar or oval distortion
strength in Run A, estimated from the maximum $Q_2$ over radius 
of the ratio: $m = 2$ tangential force divided by the radial
 force. } 
\label{bars}
\end{figure}

\begin{figure}[ht]
\rotatebox{-90}{\includegraphics[width=6cm]{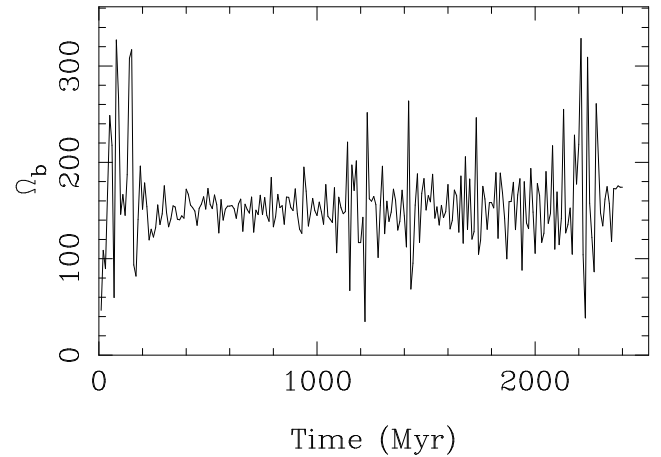}}
\caption{  Evolution in Run A of the pattern speed
of the $m=2$ distortion. $\Omega_b$ is estimated from
the Fourier transform of the potential every 10 Myr,
and in particular the evolution of the $m=2$ phase.}
\label{omeg}
\end{figure}

To explore this last solution,
and to account for only a weak oval distortion in the disk,
the disk must not be completely self-gravitating, and a significant
part of the mass should exist in the spherical bulge. This 
corresponds to the initial mass distribution of Run A 
(Table \ref{condini}). The rotation curve for this mass
distribution is plotted in Fig. \ref{vcur}. The initial gas
mass was maximum in this run, M$_{gas}$ = 4 10$^9$ M$_\odot$,
or 3\% of the total mass. The dissipation was also important,
with the collision time-scale of 5 Myr, and a collisional rebound
coefficient (or elasticity factor) of $f_{el}$= 0.35. 
A snapshot of the typical morphology between 1 and 2 Gyr
is plotted in Figures \ref{simplot} and \ref{simplot2}.
A very weak bar developed, as shown in Fig. \ref{bars}, with a
high pattern speed, above the maximum
of the $\Omega -\kappa/2$ curve (see  Fig. \ref{omeg}).
Since the mass concentrates as evolution proceeds,
the $\Omega -\kappa/2$ curve rises, and there is 
just one ILR.
The nuclear ring falls at the radius of the maximum 
of the $\Omega -\kappa/2$ curve.  For all other runs of this type
(Ai in Table \ref{condini}), the pattern speed is always high, between
100 and 150 km/s/kpc, and the ILR, the CR, and the OLR
resonances are at about 1, 2-3, and 4-5 kpc.

There is very little gas inside the nuclear ring, but still some
gas is flowing inward. There could be intermittent instabilities of
the very sharp inner edge of the ring, and some gas clouds may flow inwards. 
These are not part of a wave, but might be sheared into transient
material spiral arms through differential rotation inside in the ring.
If they have no fixed position with respect to the oval pattern,
the torques exerted on them cancel out.

\begin{figure}[ht]
\rotatebox{-90}{\includegraphics[width=6cm]{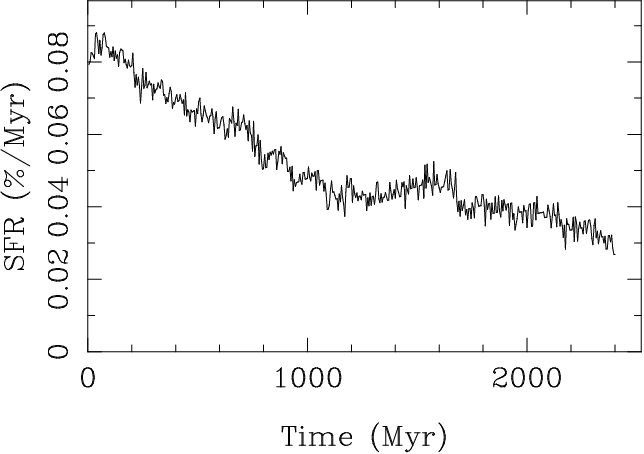}}
\caption{ Star formation rate as a function of time for Run A.
It is estimated as the percentage of the gas mass transformed into
stars per Myr.} 
\label{sfra}
\end{figure}

The evolution of the overall circumnuclear star formation rate is indicated
in Fig. \ref{sfra}. 

\begin{figure*}
\rotatebox{-90}{\includegraphics[width=5cm]{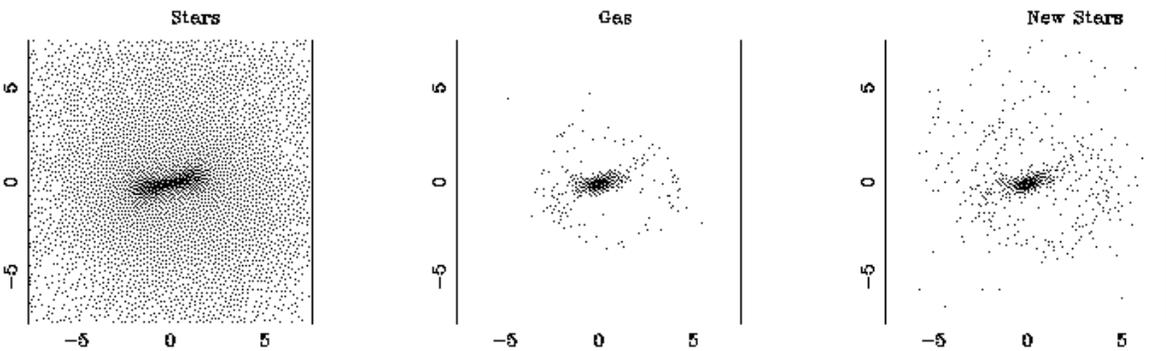}}
\caption{ {\bf Left:} Gray scale plot of the stellar component
distribution, in Run B (the gray scale is linear and axes are in kpc).
{\bf Middle:}  Gray scale plot of the corresponding gaseous component
(the gray scale is logarithmic).
{\bf Right:}  Plot of the stars formed during the simulation
(the gray scale is logarithmic).}
\label{bplot}
\end{figure*}

At the other extreme, Run B has most of the mass in its
self-gravitating stellar disk.
It is therefore highly
unstable to bar formation, although the gas mass here is only
M$_{gas}$ = 1.6 10$^9$ M$_\odot$, $\sim$ 1.5\% of the total.
The dissipation was made less important,
with the collision time-scale of 10 Myr, and a collisional rebound
coefficient of $f_{el}$=0.65. The resulting morphology between 1 and 2 Gyr,
is plotted in Figure \ref{bplot}.

In this run, the bar is so strong, and the radial gas flow
so important that only a small gas fraction remains in the 
disk at the end. Moreover, the radius of the ring shrinks, and
the ring disappears: even if the bar were destroyed later, no ring
would survive the bar. However the bar remains, since
the gas mass is insufficient to dissolve it via gas infall.

\section{Discussion}
\label{discuss}

The results of runs A and B have clarified the possible nature
of the present nuclear ring on NGC7217: the most likely solution is
that the ring has been formed by a recent weak oval distortion,
which has declined in strength but still plays a role in maintaining
the sharp inner boundary of the ring in molecular gas.
The galaxy is not likely to have been strongly barred in the recent past
(i.e. in the last 700 Myr or so),
since the ring would not have formed at its observed location with a circular
shape. The massive bulge in this
galaxy stabilizes the disk, and supports the weak bar solution.
If the galaxy were strongly barred in the past, such a configuration must date back
to prior to the building of the extended bulge.

\subsection{Other runs}
For the sake of simplicity, we have
 described two models compatible with the observed rotation curve with rather
 extreme bulge-to-disk mass ratios.
 However, many other runs with variations
of the main parameters were carried out. 
These experiments have revealed which
 parameters play the largest role in the dynamics: the disk mass ratio, the
 gas fraction, and the character of dissipation.
All three act on the stability of the disk, but
differently. When the disk mass is too dominant, a
strong bar develops, but then the gas is driven too fast toward the center,
and no ring forms. In the intermediate cases, the remaining gas
forms a resonant ring that is too elliptical in shape. When stars form out of this
gas, the stellar nuclear ring cannot be made circular enough
to comply with the observed morphology. Rings as circular as
observed are only found for a very weak bar.

The efficiency of forming a contrasted ring depends mainly on the amount
of gas and its dissipation rate. For an almost axisymmetric potential,
the gas must be self-gravitating and very cold to react as
observed. The absence of strong perturbations then maintains
the cold character of the gas component.

Even in the case of the development of a weak bar, the gas disappears
progressively because of star formation and slight inflow
toward the center; consequently the disk stabilizes and
the bar slowly fades away (Fig. \ref{bars}).  There still
remains  a strong enough perturbation to maintain a positive torque for the gas
inside the nuclear ring, and therefore to explain the ring's sharp edge.
This is necessary, since the gas distribution has a natural
tendency to smooth out any sharp structure, through its
small viscosity and velocity dispersion.

Although our best fitting model to observed geometry 
 excludes the possibility that there
has been a strong bar in the recent past (700 Myr or so), it does not mean 
that the galaxy has never been strongly barred. On the contrary, the presence
of the three rings (nuclear, inner, and outer ring) suggests the existence
of a strong bar in the more distant past. Indeed, the time-scale of
formation of the outer structures is about one order of magnitude larger
than that for the nuclear structures.  Therefore the most likely scenario is that the outer
ring was resonant with a strong bar several Gyrs ago. 
Since the present study regards the nuclear region only, the galaxy long-term
history cannot be constrained.

\subsection{Spheroidal component}

NGC\,7217 is one of
the most spheroid-dominated spirals known (see B95). 
It has not only the usual concentrated bulge, but also
 an extended, luminous, nearly spherical halo, suggesting a "bulge" even more
 extended than the disk.
Buta et al (1995) interpreted this either as the outer
regions of the bulge, or as a separate stellar halo component.
If this component is added to the central bulge to make the spheroid component,
the spheroid-to-disk luminosity ratio is 2.3, an
exceptionally large ratio.  This peculiarity might 
explain the regularity of the structures observed in NGC7217's disk.

Since the galaxy is nearly face-on however,  it is difficult to disentangle
the contribution of the spheroidal and disk components, and the
interpretation in terms of a disk or stellar halo is  still uncertain.
Also, the mass-to-light ratios of the various components are unknown,
making it useful for us to explore several spheroidal-to-disk mass
ratios. We note, however, that the observed light ratio between bulge and disk
components favors a non-dominant disk component, and therefore
also supports run A as the best solution.

\subsection{Gas accretion}

The gas content has a large influence on the dynamics of the galaxy. A large
amount of gas makes the disk more unstable to bar formation, and the 
formation of rings is only possible with gas present, the stellar nuclear ring being
subsequently created through star formation in the gas ring. An important gas
inflow toward the center driven by a strong bar can destroy the bar,
and thus the gas regulates the bar strength. 
This process means that a bar is
often a transient feature in a galaxy's life, and external accretion
is required for a galaxy to experience several bar episodes in a Hubble
time (Bournaud \& Combes 2002).
 From a statistical study of bar strength in a large sample of nearby
galaxies, Block et al (2002) estimate that a typical galaxy
must double its own mass in 10Gyr through gas accretion. 

A more complete modelling of this galaxy would therefore include
gas accretion. 
However, given the observed low gas fraction
 of NGC7217, we do not think it is currently experiencing rapid gas accretion,
 and so we have not explored models which include a nonzero accretion rate as
 a free parameter.

\section{Conclusion}

The molecular gas in the ringed LINER NGC 7217 has been mapped with high resolution
($\sim$ 150pc) inside a radius of 1.5kpc. The CO emission is confined to  a
broad nuclear ring, remarkably regular and complete. The average radius of the ring
is 900pc, and its width is about 400pc, covering both the narrow stellar ring
(at 750pc radius) and the multi-armed spiral structure observed in the optical
at slightly larger radii. The ring sometimes splits up into several lanes (barely
resolved here) that correspond
to the spiral structure, fragmented in a few dense clumps. The most striking
feature is the sharpness of the inner boundary of the ring. Inside the ring
very little emission is found, in particular a nuclear spot of about
5 10$^5$ M$_\odot$ inside a radius of 70pc. Almost perfect circular motion
is observed in the ring. 

The action of an oval distortion is
 favored to account for the sharp ring edge.  The presence of three rings in
 this galaxy already supports
the bar hypothesis, since the locations of the three rings correspond
to bar resonances, according to the rotation curve (cf B95).
However, two scenarios are conceivable,  either the observed sharp nuclear ring is the
consequence of a recent strong bar that has now faded, or a persistent
 weak oval distortion is sufficient.
The two possibilities would have different impacts on fueling of the AGN.

As the rotation curve does not unambiguously separate disk and
 spheroidal mass components, it is of legitimate interest to explore models
 with both maximal and minimal disks.  These two
extreme models were explored with N-body models taking into account the gravity
of stars and gas, as well as star formation and feedback. 
The best fit to the observations is obtained
 when only a weak bar existed in the recent past, and a large fraction of the
 original gas mass has been consumed by star formation.  The gas was more
 strongly self-gravitating in the past and formed a very contrasted ring,
 whose continued high contrast is maintained by the weak oval distortion.
 This distortion prevents a large quantity of gas from flowing
to the nucleus and accounts
for the low gas content observed there.
In the alternate hypothesis, a strong bar does not lead to the formation 
of a contrasted ring, but instead drives a lot of gas toward the center.
This would have been observed today, unless a nuclear starburst occurred.
However the presence of such a nuclear starburst is not obvious from the galaxy
 colors.

We conclude that the sharp CO ring has been built quite recently,
when the galaxy had a weak bar in its disk and a higher gas content.
The bar has weakened now into an oval distortion, and the consumption
of the gas by star formation has now also weakened its 
self-gravity, preventing efficient fueling of the AGN.

\begin{acknowledgements}
We thank the referee, Jonathan Braine, for a careful reading of the manuscript.
We are grateful to F. Gueth for having kindly
provided the SHORT-SPACE task, and to 
Melanie Krips for her help in obtaining the 30m data.
Lourdes Verdes-Montenegro has kindly sent the H$\alpha$ image.
In this work, we have used the NED and LEDA databases.
S. L. is supported by a Marie Curie Individual Fellowship contract
  HPMF-CT-20002-01734  from the European Union.
The computations in this work have been realized on the Fujitsu
NEC-SX5 of the CNRS computing center, at IDRIS. 
\end{acknowledgements}

\end{document}